\documentclass[12pt]{article} 
\usepackage[sectionbib]{natbib}
\usepackage{array,epsfig,fancyheadings,rotating}
\usepackage[]{hyperref}  
\usepackage{sectsty, secdot}
\sectionfont{\fontsize{12}{14pt plus.8pt minus .6pt}\selectfont}
\renewcommand{\theequation}{\thesection\arabic{equation}}
\subsectionfont{\fontsize{12}{14pt plus.8pt minus .6pt}\selectfont}

\usepackage{amsmath}
\usepackage{amssymb}
\usepackage{amsfonts}
\usepackage{multirow}
\usepackage{amsthm}

\usepackage[utf8]{inputenc} 
\usepackage[T1]{fontenc}    
\usepackage{url}            
\usepackage{booktabs}       
\usepackage{nicefrac}       
\usepackage{microtype}      
\usepackage{lipsum}
\usepackage{graphicx}
\usepackage{xcolor}
\usepackage{longtable}
\usepackage{booktabs} 
\usepackage{subcaption}

\graphicspath{ {./images/} }

\usepackage[ruled,vlined]{algorithm2e} 
\usepackage{bm}                        
\usepackage{algorithmic}
\usepackage{booktabs}   
\usepackage{caption}    
\usepackage{geometry}
\usepackage{graphicx} 

\newtheorem{theorem}{Theorem}
\newtheorem{assumption}{Assumption}

\theoremstyle{definition}

\newcommand{\cA}{\mathcal{A}}
\newcommand{\cP}{\mathcal{P}}
\newcommand{\cV}{\mathcal{V}}
\newcommand{\cS}{\mathcal{S}}
\newcommand{\cM}{\mathcal{M}}
\newcommand{\cL}{\mathcal{L}}
\newcommand{\Ab}{\mathbf{A}}
\newcommand{\Pb}{\mathbf{P}}

\begin{document}
	
	
\renewcommand{\baselinestretch}{2}

\markright{ \hbox{\footnotesize\rm Statistica Sinica
    }\hfill\\[-13pt]
    \hbox{\footnotesize\rm
    }\hfill }

\markboth{\hfill{\footnotesize\rm FIRSTNAME1 LASTNAME1 AND FIRSTNAME2 LASTNAME2} \hfill}
{\hfill {\footnotesize\rm FILL IN A SHORT RUNNING TITLE} \hfill}

\renewcommand{\thefootnote}{}
$\ $\par


\fontsize{12}{14pt plus.8pt minus .6pt}\selectfont \vspace{0.8pc}
\centerline{\large\bf Connection Probabilities Estimation in Multi-layer Networks }
\vspace{2pt} 
\centerline{\large\bf via Iterative Neighborhood Smoothing}
\vspace{.4cm} 
\centerline{
Dingzi Guo$^{1}$, 
Diqing Li$^{2}$\footnote{Diqing Li is the corresponding author.
All authors contributed equally to this work.
},
Jingyi Wang$^{1}$,
Wen-Xin Zhou$^{3}$} 
\vspace{.4cm} 
\centerline{\it $^1$ Shandong University \ $^2$ Zhejiang Gongshang University\ $^3$University of Illinois Chicago}
\vspace{.55cm} \fontsize{9}{11.5pt plus.8pt minus.6pt}\selectfont

	
\begin{quotation}
\noindent {\it Abstract:}
Understanding the structural mechanisms of multi-layer networks is essential for analyzing complex systems characterized by multiple interacting layers. This work studies the problem of estimating connection probabilities in multi-layer networks and introduces a new Multi-layer Iterative Connection Probability Estimation (MICE) method. The proposed approach employs an iterative framework that jointly refines inter-layer and intra-layer similarity sets by dynamically updating distance metrics derived from current probability estimates. By leveraging both layer-level and node-level neighborhood information, MICE improves estimation accuracy while preserving computational efficiency. Theoretical analysis establishes the consistency of the estimator and shows that, under mild regularity conditions, the proposed method achieves an optimal convergence rate comparable to that of an oracle estimator. Extensive simulation studies across diverse graphon structures demonstrate the superior performance of MICE relative to existing methods. Empirical evaluations using brain network data from patients with Attention-Deficit/Hyperactivity Disorder (ADHD) and global food and agricultural trade network data further illustrate the robustness and effectiveness of the method in link prediction tasks. Overall, this work provides a theoretically grounded and practically scalable framework for probabilistic modeling and inference in multi-layer network systems.

\vspace{9pt}
\noindent {\it Key words and phrases:}
Multi-layer networks, Graphon estimation, Connection probability, Neighborhood smoothing.
\par
\end{quotation}\par

\def\thefigure{\arabic{figure}}
\def\thetable{\arabic{table}}

\renewcommand{\theequation}{\thesection.\arabic{equation}}

\fontsize{12}{14pt plus.8pt minus .6pt}\selectfont

\section{Introduction}

Network analysis has become a central paradigm for studying complex systems, and the past two decades have seen rapid methodological development with applications in sociology \citep{eagle2009inferring,leskovec2010empirical}, neurobiology \citep{rahiminejad2019topological,calderer2021community}, and statistical physics \citep{boccaletti2006complex,goldenberg2010survey}. Classical network models typically describe a system using a single-layer representation, where nodes denote system components and edges encode pairwise interactions. However, many real-world systems consist of multiple interdependent layers rather than a single homogeneous structure. For example, individuals in social systems engage across multiple online platforms such as \texttt{Twitter}, \texttt{Facebook}, and \texttt{Instagram}; cities in transportation networks are linked through aviation, rail, and road infrastructures; and biological functions are jointly governed by gene regulatory, protein-protein interaction, and metabolic networks. To capture such structural complexity, multi-layer network models \citep{kivela2014multilayer} have been proposed. By explicitly modeling intra-layer interactions and inter-layer dependencies, these models provide a principled and flexible framework for representing multidimensional and heterogeneous interaction structures.

A fundamental problem in multi-layer network research is to elucidate the mechanisms that drive the formation of topological structures. As in the single-layer setting, generative modeling offers a principled statistical framework for studying such mechanisms. Within this framework, connection probabilities are central: they quantify the likelihood of edge formation between node pairs and thus encode the underlying probabilistic structure of the network. However, accurately estimating these probabilities in multi-layer networks is challenging. Effective estimation requires an adaptive strategy that borrows strength across layers, capturing correlated intra-layer connectivity patterns and inter-layer dependencies, while simultaneously accounting for layer-specific node heterogeneity and distinct generative mechanisms.

While link prediction has been extensively studied in single-layer networks, the corresponding literature for multi-layer networks remains relatively limited. Existing approaches can be broadly grouped into two categories. The first category infers potential links by computing similarity scores between non-adjacent node pairs \citep{Pham2022}. Although often effective in practice, such heuristic methods lack an explicit generative mechanism and, as a consequence, do not yield formal theoretical guarantees. The second stream consists of probabilistic network models. Although often originally motivated by community detection, these models naturally facilitate link prediction by explicitly characterizing edge formation through latent structures. Within this framework, \citet{Jing2021community} and \citet{Xu03072023} proposed tensor-based approaches that leverage Tucker decomposition to recover latent community structures by treating the multi-layer adjacency matrices as a high-order tensor. While empirically effective, their theoretical guarantees typically rely on strict low-rank assumptions that may impose excessive structural homogeneity across layers. More recently, \citet{agterberg2025joint} introduced DC-MASE, a joint spectral clustering method for multi-layer community detection that accommodates layer-specific degree-correction parameters and block connectivity matrices, but enforces a strictly common community membership across all layers. This assumption may be overly restrictive in applications where community structures are misaligned or evolve across layers. Alternatively, \citet{macdonald2022latent} proposed the MultiNeSS framework, which decomposes node latent representations into shared and layer-specific components, providing greater flexibility. However, its performance may deteriorate in highly sparse regimes, and its computational cost can pose challenges for large-scale networks.

In summary, although existing methods for link prediction and related probabilistic modeling in multi-layer networks have made notable progress, they continue to rely on strong model assumptions, exhibit limited flexibility in accommodating heterogeneity, and in some cases incur substantial computational costs. These considerations motivate the development of a connection probability estimation framework that more flexibly adapts to inter-layer heterogeneity while maintaining principled theoretical guarantees and computational efficiency.

Nonparametric graphon models provide a natural foundation for such a framework. In recent years, graphon estimation has received increasing attention in network analysis \citep{wolfe2013graphon,Gao2015graphon,Choi2017graphon,Chandna2021graphon} due to its capacity to represent complex generative mechanisms in a flexible and interpretable manner. For single-layer networks, \cite{zhang2017estimating} and \cite{ICE} proposed neighborhood smoothing estimators for piecewise Lipschitz graphon functions, establishing the optimal error rate among all computationally feasible procedures. Building on these foundations, we propose the Multi-layer Iterative Connection Probability Estimation (MICE) method, tailored to the multi-layer graphon framework \citep{he2026jointestimationedgeprobabilities}. Crucially, unlike the classic bivariate single-layer graphon, this multi-layer framework incorporates a layer-specific latent variable $\eta_k$, leading to a ternary graphon function $f(\xi_i,\xi_j,\eta_k)$ that provides a flexible way for modeling heterogeneous multi-layer networks. This framework is remarkably versatile and encompasses many complex network structures as special cases. For example, if $\eta_k$ denotes the time point $t_k$, the model reduces to a dynamic graphon 
\citep{dynamicPensky,chandna2020nonparametric}; when $\eta_k$ represents observational covariates, each layer corresponds to a covariate network \citep{chandna2020nonparametric}; and when $\eta_k$ takes only two values, the framework can be adapted for change-point detection in temporal networks \citep{wang2021optimal}. Furthermore, when $f$ is a piecewise constant function, the framework naturally recovers the multi-layer stochastic block model (MSBM) proposed by \cite{MSBM_Paul}.

The main contributions of this work are as follows.

\begin{itemize}
    \item  First, we develop a general framework for estimating connection probabilities in multi-layer networks under mild regularity conditions. In contrast to existing methods that typically impose restrictive structural assumptions, our framework is highly flexible and accommodates a broad spectrum of architectures, ranging from homogeneous to strongly heterogeneous systems. The proposed MICE method employs a dual-level smoothing strategy that adaptively integrates information from both local node neighborhoods and global layer similarities, allowing it to balance inter-layer heterogeneity with shared structural patterns in a data-driven and robust manner.

    \item Second, we introduce an iterative refinement algorithm based on a new pairwise distance metric. Unlike static, one-step neighborhood smoothing techniques, our approach dynamically updates node- and layer-level similarity measures at each iteration using the current probability estimates. This feedback mechanism progressively sharpens the identification of structurally comparable peers and effectively corrects initial estimation errors. Through adaptive neighborhood updates, MICE captures subtle and complex dependencies that static methods frequently fail to detect.

    \item  Third, we establish the theoretical consistency and convergence guarantees of the MICE estimator. Our analysis shows that incorporating multi-layer information leads to substantial accuracy gains. In particular, as the number of layers increases, the estimator achieves provably faster convergence rates than any computationally feasible single-layer method, underscoring the intrinsic statistical advantages of aggregating information across layers. Under mild regularity conditions, we further show that the proposed estimator attains the optimal convergence rate, matching that of an oracle estimator. Finally, we assess the finite-sample performance and practical utility through extensive simulations and applications to brain network data from ADHD patients and to global food and agricultural trade networks.
\end{itemize}

The remainder of this paper is organized as follows. Section~\ref{sec:method} presents the model formulation, methodological framework, and algorithmic components. Section~\ref{sec:theory} develops the theoretical guarantees of the proposed method. Sections~\ref{sec:sim} and~\ref{sec:real} report simulation studies and empirical applications that illustrate the finite-sample performance of MICE. Section~\ref{sec:conclusion} concludes with a brief summary and discussion. Additional simulation results, complete proofs of the main theorems, and auxiliary lemmas are provided in the Supplementary Materials.

\section{Methodology}
\label{sec:method}

\subsection{Notation and model}

Consider a multi-layer network with $n$ nodes and $K$ layers, represented by an adjacency tensor $\cA\in\{0,1\}^{n\times n\times K}$, where $\cA_{ijk} = 1$ indicates the presence of an edge between nodes $i$ and $j$ in layer $k$, and $\cA_{ijk} = 0$ otherwise. Let $\cV=[n]$ and $\cL=[K]$ denote the node set and the layer set, respectively. Throughout this paper, we focus on undirected networks without self-loops; accordingly, for each layer $k\in [K]$, the $k$-th adjacency matrix slice, denoted by $\Ab^k = \cA_{\cdot \cdot k}$, is symmetric with zero diagonal entries. Assume that each layer is generated from an exchangeable random graph model. By the Aldous-Hoover representation theorem \citep{aldous1981representations,hoover1979relations}, there exist node-specific latent variables $\xi_i$ and a layer-specific graphon function $f_k$ such that the edge probability satisfies $\cP_{ijk}=f_k(\xi_i, \xi_j)$. To capture dependence across layers in a coherent manner, we further introduce a layer-specific latent variable $\eta_k$ representing the latent position of layer $k$. This yields the multi-layer graphon model 
$\cA_{ijk}\sim\text{Bernoulli}(\cP_{ijk}), \cP_{ijk}=f(\xi_i,\xi_j,\eta_k)$, where $f(\cdot)$ is the multi-layer graphon function \citep{chandna2020nonparametric,he2026jointestimationedgeprobabilities}. This formulation encompasses a wide range of multi-layer architectures, from homogeneous layers sharing a common structure to highly heterogeneous systems. Let $\cP\in[0,1]^{n\times n\times K}$ denote the full tensor of connection probabilities, and let $\Pb^k = \cP_{\cdot \cdot k}$ be the probability matrix for layer $k$.

In this paper, we aim to estimate the connection probability tensor $\cP$ (or equivalently, the probability matrices $\Pb^k$ for each layer $k\in[K]$) based on the observed adjacency tensor $\cA$. Because each entry $\cA_{ijk}$ provides only a single Bernoulli realization of the underlying probability $\cP_{ijk}$, direct likelihood-based estimation is ill-posed without additional structural assumptions. To overcome this difficulty, we exploit the smoothness inherent in the multi-layer graphon model, which permits the sharing of information across similar nodes and similar layers and thus enables the construction of consistent estimators.

Specifically, when the latent position $\eta_k$ of layer $k$ is close to that of another layer $\eta_{k'}$, the smoothness of the graphon function $f$ implies that the corresponding connection probability matrices $\cP_{\cdot\cdot k}$ and $\cP_{\cdot\cdot k'}$ must be similar. This observation naturally motivates grouping structurally comparable layers into sets of \textit{neighboring layers}, denoted by $\mathcal{S}^k = \{k':\cP_{\cdot \cdot k}\approx \cP_{\cdot \cdot k'}\}$. Likewise, within a fixed layer $k$, if the latent position $\xi_{i'}$ of node $i'$ is close to that of node $i$, the smoothness of $f$ in its first argument ensures that $f(\xi_{i'},\cdot,\eta_k)\approx f(\xi_{i},\cdot,\eta_k)$, and thus $\cP_{i\cdot k} \approx \cP_{i'\cdot k}$. In this case, the observed edges $\cA_{i'jk}$ serve as proxy observations for $\cP_{ijk}$. This leads to groups of nodes exhibiting similar connectivity patterns, referred to as \textit{neighboring nodes}. For any layer $k'\in\mathcal{S}^k$, we denote by $\mathcal{S}_i^{k'}=\{i':\cP_{i\cdot k'}\approx \cP_{i'\cdot k'}\}$ the set of nodes in layer $k'$ that are analogous to node $i$.

Based on these notions, the connection probability $\cP_{ijk}$ can be estimated by aggregating information from neighboring nodes and neighboring layers, leading to the oracle estimator
\begin{equation}
    \widetilde{\cP}_{ijk}=\widetilde{\cP}_{jik}=\frac{\sum_{k' \in \mathcal{S}^k} \sum_{i' \in {\mathcal{S}_i^{k'}}} \sum_{j' \in {\mathcal{S}_j^{k'}}}\cA_{i'j'k'}}{\sum_{k' \in \mathcal{S}^k} |\mathcal{S}^{k'}_i| |\mathcal{S}^{k'}_j|}=(s_i^*s_j^*t_k^*)^{-1}\sum_{k' \in \mathcal{S}^k} \sum_{i' \in {\mathcal{S}_i^{k'}}} \sum_{j' \in {\mathcal{S}_j^{k'}}}\cA_{i'j'k'}.
    \label{eq:oracle}
\end{equation}
Because the network is symmetric (i.e., $\widetilde{\mathcal{P}}_{ijk} = \widetilde{\mathcal{P}}_{jik}$), we employ a uniform smoothing bandwidth. In particular, we assume that neighborhood sizes are stable across structurally similar layers and denote $s_i^* = |\mathcal{S}_i^{k'}|$ and $t_k^* = |\mathcal{S}^k|$. The construction of these neighboring sets is described in the next subsection.

\subsection{Iterative connection probability estimation for multi-layer networks}
\label{subsec:MICE_alg}

We first consider the ideal oracle setting, in which the true link probability tensor $\cP$ is known. In this scenario, neighboring nodes and layers can be identified directly through population-level distance measures defined on the probability matrices. Specifically, the distance between two layers $k$ and $k'$ is measured by the normalized squared Frobenius distance $d_{kk'}=n^{-2}\lVert \Pb^k-\Pb^{k'}\rVert_F^2$. Likewise, within a given layer $k$, the distance between nodes $i$ and $i'$ is quantified by the normalized squared Euclidean distance between their corresponding
probability vectors, $d_{ii'}^{k}=n^{-1}\lVert \Pb^k_{i\cdot}-\Pb^k_{i'\cdot}\rVert_2^2$. These distance metrics provide a principled basis for defining neighboring layers and neighboring nodes in the oracle regime.

Based on these distance metrics, the neighboring layer set for layer $k$ is defined as $\mathcal{S}^k=\{k': 0 \leq d_{kk'}< d_{t_k^*}\}$, where $d_{t_k^*}$ denotes the $t_k^*$-th smallest value among $\{ d_{kk'} : k' \neq k \}$. Thus, $\mathcal{S}^k$ contains layer $k$ together with the $t_k^*-1$ layers whose probability matrices are closest to $\Pb^k$. For each $k'\in \mathcal{S}^k$, the neighboring node set for node $i$ is then constructed as $\mathcal{S}^{k'}_i=\{i':0<d_{ii'}^{k'}\leq d_{s_i^*}\}$, where $d_{s_i^*}$ is the $s_i^*$-th smallest element of the intra-layer distance set $\{d_{ii'}^{k'} : i' \neq i\}$. In this way, $\mathcal{S}^{k'}_i$ collects $s_i^*$ nodes in layer $k'$ whose connectivity profiles most closely resemble that of node $i$.

In practice, however, only the adjacency tensor $\cA$ is observed, while the underlying connection probabilities $\cP$ are unknown. Consequently, the population-level distances and the oracle neighborhoods defined above cannot be computed directly. To address this challenge, the proposed MICE method employs an iterative refinement strategy. Starting from an initial estimate $\widehat{\Pb}^k$, we construct plug-in estimators of the inter-layer and intra-layer distances, which yield the estimated neighboring layer sets $\widehat{\mathcal{S}}^k$ and the estimated neighboring node sets $\widehat{\mathcal{S}}^{k'}_i$ and $\widehat{\mathcal{S}}^{k'}_j$. Using these estimated neighborhoods, the connection probability estimates are updated through the smoothing operation
\begin{equation}
    \widehat{\cP}_{ijk}=\widehat{\cP}_{jik}=\frac{\sum_{k' \in \mathcal{\widehat{S}}^k} \sum_{i' \in \widehat{\cS}_i^{k'}} \sum_{j' \in {\mathcal{\widehat{S}}_j^{k'}}}\cA_{i'j'k'}}{\sum_{k' \in \mathcal{\widehat{S}}^k} |\mathcal{\widehat{S}}^{k'}_i| |\mathcal{\widehat{S}}^{k'}_j|}=(s_is_jt_k)^{-1}\sum_{k' \in \mathcal{\widehat{S}}^k} \sum_{i' \in \widehat{\cS}_i^{k'}} \sum_{j' \in {\mathcal{\widehat{S}}_j^{k'}}}\cA_{i'j'k'},
\end{equation}
where $s_i=|\widehat{\mathcal{S}}^{k'}_i|$ and $t_k=|\widehat{\mathcal{S}}^k|$ denote the cardinalities of the estimated node and layer neighborhoods. Based on the updated estimate, we recompute the pairwise distances between nodes and layers, which refines the neighborhood construction and, in turn, produces an improved estimator $\widehat{\Pb}^k$. This iterative procedure, alternating between refinement and probability smoothing, is repeated until convergence. The complete computational scheme is summarized in Algorithm~\ref{alg}.

To fully implement the proposed algorithm, two practical considerations require specification: the choice of neighborhood sizes and the initialization strategy. For the neighborhood sizes, we follow the optimal smoothing theory for single-layer graphons \citep{zhang2017estimating}. Accordingly, we set the node-level neighborhood size as $s_i = D_i (n\log n)^{1/2}$. Extending this principle to the multi-layer setting, we specify the layer-level neighborhood size as $t_k = G_k (K\log K)^{1/2}$. The constants $D_i$ and $G_k$ are further examined through a comprehensive sensitivity analysis in Section~\ref{sec_simu_para}, which shows that the optimal configuration varies subtly with the network's structural complexity. Based on these findings, we report the recommended values and provide their justifications. The second consideration concerns the initialization $\widehat{\mathbf{P}}^{k(0)}$. While any computationally feasible estimator that consistently approximates the underlying probability tensor is, in principle, acceptable, a high-quality initialization can substantially accelerate convergence. In this work, we adopt the MNS estimator of \citet{he2026jointestimationedgeprobabilities} as a warm-start for the iterative procedure.

\begin{algorithm}[htbp]
    \caption{Multi-layer iterative connection probability estimation.}
    \label{alg}
    \KwIn{observed adjacency matrices $\Ab^k,k=1,\cdots,K$; initial connection probability estimators $\widehat{\mathbf{P}}^{k(0)},k=1,\cdots,K$; neighborhood sizes $s_i,s_j$ and $t_k$; threshold $\delta_0 > 0$.}
    \KwOut{connection probability estimators $\widehat{\mathbf{P}}^k, k=1,\cdots,K$.}
    
    Let $\delta_{\Pb} = +\infty$ and $m = 0$\;
    \While{$\delta_{\Pb} > \delta_0$}{
        For each layer pair $(k, k')$ and node pair $(i, i')$, compute the distances $\hat{d}_{kk'}=\|\widehat{\Pb}^{k(m)}-\widehat{\Pb}^{k'(m)}\|^2_F/n^2$  and $\hat{d}_{ii'}^k = \|\widehat{\Pb}^{k(m)}_{i\cdot} - \widehat{\Pb}^{k(m)}_{i'\cdot}\|_2^2 / n$.\\
        
        For each layer $k\in \cL$ and node $i\in \cV$, construct the estimated neighboring layers $\widehat{\mathcal{S}}^k=\{k':0\leq \hat{d}_{kk'}< d_{{t}_k}\}$ and neighboring nodes  $\widehat{\mathcal{S}}^{k'}_i=\{i':0<\hat{d}_{ii'}^{k'}\leq d_{s_i}, k'\in\widehat{\cS}^k\}$.\\
        
        For each node pair $i, j\in \cV$ within layer $k$, update the connection probability estimator:
        \(
        \widehat{\cP}_{ijk}^{(m+1)}=\widehat{\cP}_{jik}^{(m+1)}=(s_is_jt_k)^{-1}\sum_{k' \in \widehat{\cS}^k} \sum_{i' \in \widehat{\cS}_i^{k'}} \sum_{j' \in {\widehat{\cS}_j^{k'}}}\cA_{i'j'k'}.
        \)
        
        Set $\widehat{\Pb}^{k(m+1)} = \widehat{\cP}_{\cdot \cdot k}^{(m+1)}$ for all $k=1,\ldots, K$, and update $\delta_{\Pb} = \frac{\sum\limits_{k=1}^K\sqrt{\|\widehat{\Pb}^{k(m+1)} - \widehat{\Pb}^{k(m)}\|_F^2}}{\sum\limits_{k=1}^K\sqrt{\|\widehat{\Pb}^{k(m)}\|_F^2}}$.

        Let $m = m + 1$.
    }
    \Return{$\widehat{\mathbf{P}}^k = \widehat{\mathbf{P}}^{k(m)},k=1,\cdots,K$.}
\end{algorithm}

\section{Theoretical Properties}
\label{sec:theory}

In this section, we establish the consistency of the MICE estimator. The analysis relies on a set of assumptions that are standard in the graphon literature.

\begin{assumption}\label{asmp:lipschitz}
    There exist partitions of $[0,1]$, denoted by $\mathcal{I} = \{I_r\}_{r=1}^R$ and $\mathcal{J} = \{J_t\}_{t=1}^T$, where $I_r = [x_{r-1}, x_r)$ and $J_t = [z_{t-1}, z_t)$. Within this framework, the multi-layer graphon $f: [0,1]^3 \to [0,1]$ is piecewise Lipschitz on the grid cells $\mathcal{I} \times \mathcal{I} \times \mathcal{J}$. More precisely, there exist positive constants $L_n$ and $L_K$ such that, for any grid cell $I_r \times I_s \times J_t$, the following Lipschitz conditions hold: $|f(u_1, v, w) - f(u_2, v, w)| \leq L_n |u_1 - u_2|$,
$|f(u, v_1, w) - f(u, v_2, w)| \leq L_n |v_1 - v_2|$, and $|f(u, v, w_1) - f(u, v, w_2)| \leq L_K |w_1 - w_2|$, for all $u, u_1, u_2 \in I_r$, $v, v_1, v_2 \in I_s$, and $w, w_1, w_2 \in J_t$.
\end{assumption}

Assumption \ref{asmp:lipschitz} formalizes a piecewise Lipschitz condition on the multi-layer graphon, following the framework introduced in \citet{he2026jointestimationedgeprobabilities}. This requirement is considerably weaker than the structural assumptions typically imposed in classical graphon estimation, such as monotonicity \citep{Bickel2009ANV} or global Lipschitz continuity \citep{pmlr-v32-chan14}. Those stronger conditions often exclude many practically relevant and interpretable models, including stochastic block models with piecewise constant connection probabilities. By contrast, Assumption \ref{asmp:lipschitz} accommodates a wide range of multi-layer network structures, including multi-layer stochastic block models \citep{MSBM_Paul,Jing2021community} and dynamic graphon models \citep{zhao2019change,dynamicPensky}. This flexibility ensures that our theoretical guarantees apply to a broad class of complex real-world networks.

\begin{assumption}\label{asmp:dense}
    The number of partitions $R$ increases with $n$ in such a way that $\min_r|I_r|/(n^{-1}\log n)^{1/2}\rightarrow \infty$, and the number of partitions $T$ increases with $K$ such that $\min_t |J_t|/(K^{-1}\log K)^{1/2}\rightarrow \infty$. Here $|I_r|$ and $|J_t|$ denote the lengths of the intervals $I_r$ and $J_t$, respectively.
\end{assumption}

For any latent positions $\xi_i, \eta_k \in [0,1]$, let $I(\xi_i)$ and $J(\eta_k)$ denote the partition intervals that contain $\xi_i$ and $\eta_k$, respectively. We define the restricted local neighborhoods by $\mathcal{N}_i(\Delta_n) = [\xi_i - \Delta_n,  \xi_i + \Delta_n] \cap I(\xi_i)$ and $\mathcal{M}_k(\Delta_K) = [\eta_k - \Delta_K, \eta_k + \Delta_K] \cap J(\eta_k)$. By construction, the graphon $f(u,v,w)$ satisfies a piecewise Lipschitz condition in $(u,v)$ on $\mathcal{N}_i(\Delta_n)$ (for fixed $w\in J_t$) and in $w$ on $\mathcal{M}_k(\Delta_K)$ (for fixed $u,v\in I_r \times I_s$). Adapting Lemma 1 of \citet{zhang2017estimating} to the multi-layer setting, we specify the neighborhood radii as $\Delta_n = \Big[C + (C_1+4)^{1/2}\Big](n^{-1}\log n)^{1/2}$ and $\Delta_K = \Big[C' + (C_1'+4)^{1/2}\Big](K^{-1}\log K)^{1/2}$, for constants $C, C_1, C', C_1' > 0$. Under these choices, with probability at least $1 - 2n^{-C_1/4}-2K^{-C_1'/4}$, we have simultaneously that
$\min_{i \in \cV} \big|\{ i' : \xi_{i'} \in \mathcal{N}_i(\Delta_n)\} \big| \geq C (n\log n)^{1/2}$ and $\min_{k \in \cL} \big|\{ k' :\eta_{k'} \in \mathcal{M}_k(\Delta_K)\} \big| \geq C (K\log K)^{1/2}$. This result ensures that, for sufficiently large $n$ and $K$, the number of similar nodes and layers available for smoothing grows at least on the order of $(n\log n)^{1/2}$ and $(K\log K)^{1/2}$ with high probability.

If the true neighboring sets  $\mathcal{S}^k$ and $\mathcal{S}_i^{k'}$ were known in advance, the oracle estimator $\widetilde{\cP}$ defined in \eqref{eq:oracle} could be computed directly from the observed adjacency tensor $\cA$. This oracle estimator provides a theoretical benchmark: it reflects the best performance that any neighborhood smoothing method can achieve under perfect knowledge of the underlying neighborhoods. The next theorem establishes an upper bound on its estimation error.

\begin{theorem}\label{theorem:oracle}
    Let $C_l,C_m,C_l',C_m',\tilde{C}_q>0$ be some sufficiently large constants, where $l\in\{2,3\}$, $m\in\{4,5,6,7,8\}$, and $q\in \{1,2,3\}$. Define $C_*=\max\{C_l/3,C_m\}$ and $C_*'=\max\{C_l'/3,C_m'\}$. Then, with probability at least $1-2n^{-C_1/4}-2K^{-C_1'/4}-14n^{-C_*'}K^{-C_*}$, the following error bound holds for any layer $k\in [K]$:
    $$
    \frac{\|\widetilde{\Pb}^k-\Pb^k\|^2_F}{n^2}\leq \tilde{C}_1\frac{\log n}{n}+\tilde{C}_2\frac{\log K}{K}+\tilde{C}_3\frac{\sqrt{2\log n+\log K}}{[n(\log n)K(\log K)]^\frac12}.
    $$
\end{theorem}

Theorem \ref{theorem:oracle} establishes the theoretical benchmark for estimation error under oracle knowledge of the neighborhoods. The result explicitly characterizes how the error decreases as $n$ and $K$ grow. In particular, when $K$ is sufficiently large, the convergence rate improves upon several existing results: the $O((\log n/n)^{1/2})$ rate for single-layer neighborhood smoothing \citep{zhang2017estimating}, the $O((n^{-3}\log n)^{1/4})$ rate for single-layer iterative estimation \citep{ICE}, and the $O((\log n/nK)^{1/3})$ rate for standard multi-layer smoothing \citep{he2026jointestimationedgeprobabilities}. Moreover, when $K \gtrsim O(n)$, the oracle rate approaches $O(\log(n) /n)$, matching the minimax lower bound for the estimation of an individual layer derived in \cite{Gao2015graphon}. These comparisons highlight the statistical benefits of leveraging information from structurally similar layers to improve estimation accuracy.

However, the validity of the neighborhood smoothing principle requires that the node and layer dimensions $n$ and $K$ be sufficiently large so that the nearest neighbors selected by the quantile thresholds are genuinely close in structure. When either $n$ or $K$ is too small, the identified neighbors may in fact lie far from the target node or layer, potentially inducing non-negligible smoothing bias. In particular, a moderate number of layers is needed to ensure the theoretical consistency of inter-layer smoothing. Section \ref{sec:sim} evaluates MICE across a variety of configurations with different values of $K$, and the results show that the method delivers strong performance in most settings. Notably, when $K$ is limited, the framework naturally reduces to a single-layer procedure by restricting the layer neighborhood to $\mathcal{S}^k=\{k\}$, in which case MICE becomes the standard ICE method \citep{ICE} that relies solely on intra-layer information.

In practical settings, exact recovery of the true neighborhood structures is rarely achievable. Theorem \ref{theorem:robust} shows that our method is robust in this regard, demonstrating that the optimal oracle error rate is retained even when the estimated neighborhoods include a small proportion of incorrectly selected members.

\begin{theorem}\label{theorem:robust}
    Assume that $\cP_{ijk}\in [a,b]$ for all $(i,j,k)\in \cV\times \cV\times \cL$, where $0<a<b<1$, and let $\widehat{\mathcal{S}}$ denote the estimated neighborhood configuration. Suppose that the number of incorrectly selected neighbors is bounded by
    $$
    \max_{(i,j,k)}\sum_{k'\in \widehat{\cS}^k}\sum_{i'\in \widehat{\cS}_i^{k'}}\sum_{j'\in \widehat{\cS}_j^{k'}}\mathbb{I}((i',j',k')\notin\mathcal{S}_i^{k'}\times\mathcal{S}_j^{k'}\times\mathcal{S}^k)\le\ e(n,K)
    ,$$ 
    where the tolerance threshold $e(n, K)$ is given by
    \begin{equation*}
        \begin{split}
            e(n,K):=&\frac{\sqrt{C_9}}{b-a}\biggl(\max\biggl[\left\{n(\log n)^{3}K\log K\right\}^\frac12, n(\log n)(\log K),\\
            &(2\log n+\log K)^{\frac14}(n^3(\log n)^3K(\log K))^{\frac14}\biggr]\biggr).
        \end{split}
    \end{equation*}
    Then, there exist positive constants $\tilde{C}'_1$, $\tilde{C}'_2$, and $\tilde{C}'_3$ such that, with probability at least $1-2n^{-C_1/4}-2K^{-C_1'/4}-14n^{-C_*'}K^{-C_*}$, the estimator $\widehat{\mathbf{P}}^{k}$ constructed from $\widehat{\mathcal{S}}$ satisfies the same error bound as the oracle estimator for every layer $k$:
    $$
    \frac{\|\widehat{\Pb}^k-\Pb^k\|^2_F}{n^2}\leq \tilde{C}'_1\frac{\log n}{n}+\tilde{C}'_2\frac{\log K}{K}+\tilde{C}'_3\frac{\sqrt{2\log n+\log K}}{[n(\log n)K(\log K)]^\frac12}. 
    $$ 
\end{theorem}

A direct consequence of Theorem \ref{theorem:robust} concerns the allowable tolerance for neighborhood selection errors. Since the cardinality of the estimated neighborhood product set $|\widehat{\mathcal{S}}_i^{k'} \times \widehat{\mathcal{S}}_j^{k'} \times \widehat{\mathcal{S}}^k|$ is approximately $s_i s_j t_k$, the permissible fraction of incorrectly selected neighbors is given by
\begin{equation*}
        \frac{e(n,K)}{s_is_jt_k} \asymp \frac{\sqrt{C_9}}{b-a}\cdot\max\left[
        (\log n/n)^{\frac12},(\log K/K)^{\frac12},(2\log n+\log K)^{\frac14}(n\log nK\log K)^{\frac14}
        \right].
\end{equation*}
This result implies that as long as the proportion of misidentified neighbors remains below this threshold, the estimator continues to achieve the optimal convergence rate.


Building on this insight, the estimation accuracy can be further improved when the algorithm is initialized with a warm-start estimator that already satisfies a preliminary consistency bound. Formally, let $\widehat{\mathbf{P}}^{k(0)}$ be an initial estimator satisfying the uniform error bound
$$
    \max\limits_{i\in \cV,\;k\in\mathcal{L}}\frac{1}{n}\sum_{j=1}^n(\widehat{\cP}_{ijk}^{(0)}-\cP_{ijk})^2\leq C_{10}E(n,K),
$$
with probability at least $1-n^{-C_{11}}$. The error rate $E(n,K)$ is assumed to satisfy
$$
    \lim\limits_{n,K\to\infty}E(n,K)\Big/\left(\frac{\log n}{n}+\frac{\log K}{K}+\frac{\sqrt{2\log n+\log K}}{[n(\log n)K(\log K)]^\frac12}\right)\to \infty
$$
Under these conditions, the next theorem establishes that the proposed iterative procedure yields a strictly improved estimator relative to the warm start.

\begin{theorem}\label{theorem:final}
    Define the minimum signal separation gap $\kappa(n,K)$ as the smallest difference in connection probabilities between any true neighbor and any non-neighbor: 
    $$
    \kappa(n,K)=\min_{\substack{i,j\in \cV, i''\notin \mathcal{S}_i^{k'}\\k\in \mathcal{L}, k'\in \mathcal{S}^k, k''\notin \mathcal{S}^k}}|\cP_{ijk}-\cP_{i''jk''}|.
    $$ 
    Suppose the initial estimator $\widehat{\mathbf{P}}^{k(0)}$ satisfies the error bound $E(n,K)$ defined previously. If the separation gap is sufficiently large, in particular if 
    $$\kappa^2(n,K) \geq 8L_n^2\left[C+(C_1+4)^{1/2}\right]^2\frac{\log n}{n}+8L_K^2\left[C'+(C_1'+4)^{1/2}\right]^2\frac{\log K}{K} + 20 C_{10} E(n,K),$$
    then, using $\widehat{\mathbf{P}}^{k(0)}$ as the initialization, with probability at least $1 - 2n^{-C_1/4} - 2K^{-C_1'/4} - 14n^{-C_*'}K^{-C_*} - n^{-C_{11}}$, the updated estimator $\widehat{\mathbf{P}}_{\text{new}}^k$ attains the oracle convergence rate:
    $$
    \frac{\|\widehat{\Pb}^k_{new}-\Pb^k\|^2_F}{n^2}\leq\tilde{C}_1\frac{\log n}{n}+\tilde{C}_2\frac{\log K}{K}+\tilde{C}_3\frac{\sqrt{2\log n+\log K}}{[n(\log n)K(\log K)]^\frac12}.
    $$	
\end{theorem}

We emphasize that the separation requirement on $\kappa(n,K)$ is mild. Under the piecewise Lipschitz conditions imposed along both the node and layer dimensions, the minimal distinguishable scale is determined by local deviations in either index. In an ideal noiseless setting, the separation constant therefore satisfies $\kappa(n,K) \asymp \max\bigl(L_n\Delta_n, \, L_K\Delta_K\bigr)$, which reflects the intrinsic structural gap between a true neighbor and a non-neighbor in the node or layer direction.

In practice, however, neighborhood selection is carried out using the initial estimator $\widehat{\mathbf{P}}^{k(0)}$ rather than the true probability tensor $\cP$. Because the initialization error $E(n,K)$ generally exceeds the intrinsic structural variation in finite samples, the distances based on $\widehat{\mathbf{P}}^{k(0)}$ tend to be inflated relative to those defined on $\cP$. As a result, the theoretical analysis requires a larger lower bound on $\kappa(n, K)$ to ensure stable neighborhood recovery in this noisy regime. This condition is nevertheless non-restrictive: both $\Delta_n$ and $\Delta_K$ are allowed to converge to zero as $n, K \to \infty$. In this asymptotic setting, the required separation constant $\kappa(n,K)$ approaches its ideal scale $\max(L_n\Delta_n, L_K\Delta_K)$, matching our intuition.

In summary, once the method is initialized with a consistent warm-start estimator from an existing procedure, the iterative algorithm progressively refines the local neighborhood structures. Ultimately, the estimator achieves a convergence rate that coincides with the oracle benchmark, as if the true neighboring layers and nodes were known in advance.

\section{Simulation Studies}
\label{sec:sim}

\subsection{Algorithm Effectiveness}

In this section, we compare the finite-sample performance of MICE with four benchmark algorithms: the Iterative Connecting Probability Estimation (ICE) method \citep{ICE}, the Multiplex Networks with Shared Structure Approximation (MultiNeSS) model \citep{macdonald2022latent}, the Degree-Corrected Multiple Adjacency Spectral Embedding (DC-MASE) method \citep{agterberg2025joint}, and the Multi-Layer Neighborhood Smoothing (MNS) algorithm \citep{he2026jointestimationedgeprobabilities}.

Among these competitors, ICE is designed for single-layer networks, whereas the remaining three methods leverage multi-layer information. An important distinction is that MultiNeSS and DC-MASE were originally developed for latent position modeling and community detection, respectively, rather than for direct estimation of connection probabilities. To facilitate a fair comparison, we reconstruct the implied connection probabilities from their fitted model parameters. For MultiNeSS, the procedure returns low-rank approximation matrices $\widehat{\mathbf{F}}$ and $\widehat{\mathbf{G}}_k$, where $\widehat{\mathbf{F}}$ captures the shared latent structure across layers and $\widehat{\mathbf{G}}_k$ encodes layer-specific deviations. Following Remark 3 of Theorem 1 in \citet{macdonald2022latent}, we obtain the estimated connection probability between nodes $i$ and $j$ in layer $k$ via the additive form $\widehat{\mathcal{P}}_{ijk} = \widehat{\mathbf{F}}_{ij} + \widehat{\mathbf{G}}_{k,ij}$. Similarly, DC-MASE is based on a multi-layer Degree-Corrected Stochastic Block Model (DCSBM), under which the expected adjacency matrix for layer $k$ is $\mathbf{P}^{k} = \mathbf{\Theta}^{k} \mathbf{Z} \mathbf{B}^{k} \mathbf{Z}^\top \mathbf{\Theta}^{k}$, where $\mathbf{\Theta}^{k}$ denotes the layer-specific degree correction matrix, $\mathbf{Z}$ is the shared community membership matrix, and $\mathbf{B}^{k}$ is the block connectivity matrix. Once these parameters are estimated, we compute the corresponding connection probability matrix $\widehat{\mathbf{P}}^{k}$ directly from the DCSBM formulation.

\begin{figure}[htbp]
    \centering
    \includegraphics[width=1\textwidth]{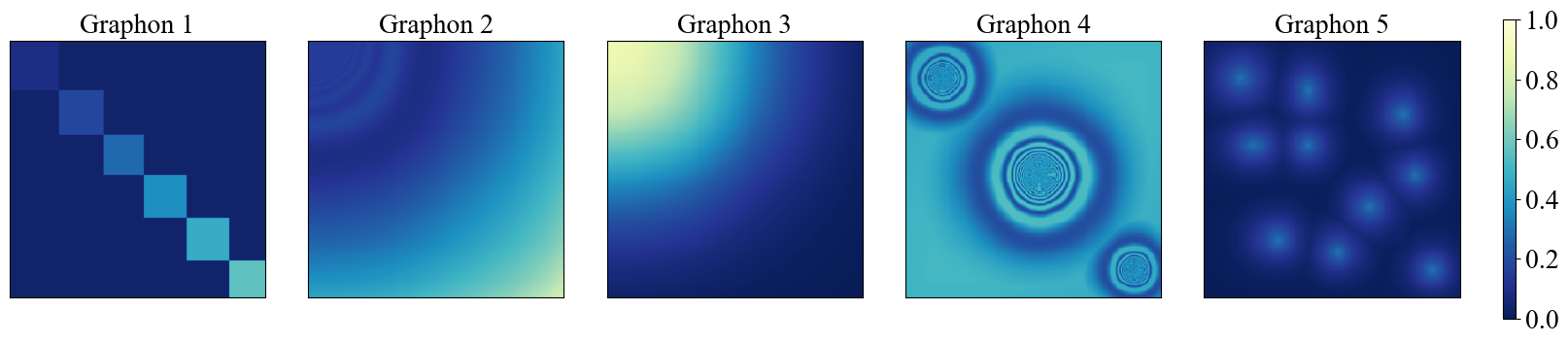}
    \caption{Connection probability matrices corresponding to Graphons 1--5.} 
    \label{fig:graphon} 
\end{figure}

To assess the robustness and broad applicability of our method, we generate multi-layer networks from five distinct graphon functions. Graphon 1 has a block structure, Graphon 2 exhibits oscillatory behavior, and Graphon 3 is full-rank with degree monotonicity. Graphons 4 and 5 are also full-rank but display more intricate structure at multiple scales. For illustration, Figure \ref{fig:graphon} shows representative heatmaps of the true connection probability matrices (the 50th layer of a network with $K=100$ layers). The explicit formulas for all five graphons are provided in Section S1 in the Supplementary Materials.

We evaluate the performance of the proposed estimator under two simulation configurations.

\begin{itemize}
\item \textbf{Scenario 1 (Varying Node Size):} We fix the number of layers at $K=100$ and vary the node size $n \in \{100, 200, 500, 1000\}$ to examine the estimator's asymptotic behavior with respect to $n$.

\item \textbf{Scenario 2 (Varying Layer Size):} We fix the node size at $n=200$ and vary the number of layers $K \in \{20, 50, 100, 200\}$ to study how performance improves with increased cross-layer information.
\end{itemize}

The average Root Mean Squared Error (RMSE) values, computed over 100 independent replications, are reported in Tables \ref{table:RMSE_node} and \ref{table:RMSE_layer}. The Mean Absolute Error (MAE) results show similar patterns and are included in Section S2 of the Supplementary Materials.

\begin{table}[h]
    \centering
    \caption{RMSE (×100) across graphons and node sizes for different methods on a fixed 100-layer network. Standard errors are reported in parentheses, and boldface indicates the best performance.
    }
    \scalebox{0.8}{
        \begin{tabular}{ccccccc}
            \toprule
            Graphon & Node Size($n$) & ICE & DC-MASE & MultiNeSS & MNS & MICE \\
            \midrule
            \multirow{4}{*}{Graphon 1} 
            & 100  & 6.43 (0.03) & 6.07 (0.11) & 3.45 (0.28) & 3.35 (0.13) & \textbf{1.74 (0.22)} \\
            & 200  & 4.39 (0.01) & 4.13 (0.07) & 2.91 (0.19) & 2.57 (0.08) & \textbf{1.24 (0.14)} \\
            & 500  & 2.44 (0.00) & 2.55 (0.03) & 2.35 (0.17) & 1.71 (0.04) & \textbf{0.89 (0.03)} \\
            & 1000 & 1.57 (0.00) & 1.80 (0.02) & 1.99 (0.11) & 1.19 (0.02) &\textbf{ 0.80 (0.03)} \\
            \midrule
            \multirow{4}{*}{Graphon 2}
            & 100  & 6.90 (0.03) & 7.40 (0.12) & 5.30 (0.34) & 4.01 (0.12) & \textbf{2.78 (0.11)} \\
            & 200  & 5.53 (0.00) & 5.58 (0.11) & 4.57 (0.19) & 3.26 (0.08) & \textbf{2.13 (0.06)} \\
            & 500  & 4.03 (0.00) & 3.99 (0.11) & 3.50 (0.12) & 2.54 (0.06) & \textbf{1.59 (0.05)} \\
            & 1000 & 2.96 (0.04) & 3.10 (0.07) & 2.81 (0.09) & 2.05 (0.04) & \textbf{1.33 (0.05)} \\
            \midrule
            \multirow{4}{*}{Graphon 3}
            & 100  & 6.92 (0.01) & 6.03 (0.15) & 4.49 (0.23) & 3.15 (0.08) & \textbf{2.36 (0.14)} \\
            & 200  & 5.71 (0.01) & 4.23 (0.07) & 4.16 (0.21) & 2.60 (0.05) & \textbf{1.72 (0.09)} \\
            & 500  & 3.44 (0.01) & 2.64 (0.03) & 3.96 (0.21) & 2.01 (0.02) & \textbf{1.22 (0.06)} \\
            & 1000 & 2.34 (0.01) & 1.86 (0.02) & 3.85 (0.21) & 1.62 (0.02) & \textbf{1.02 (0.05)} \\
            \midrule
            \multirow{4}{*}{Graphon 4}
            & 100  & 10.62 (0.01) & 10.01 (0.20) & 9.62 (0.28) & 8.02 (0.20) & \textbf{6.53 (0.34)} \\
            & 200  & 9.47 (0.01) & 8.37 (0.17) & 9.13 (0.22) & 7.51 (0.15) & \textbf{5.79 (0.29)} \\
            & 500  & 8.60 (0.01) & 7.43 (0.15) & 7.98 (0.19) & 6.80 (0.13) &\textbf{5.18 (0.12)} \\
            & 1000 & 6.86 (0.01) & 7.08 (0.14) & 6.99 (0.16) & 6.19 (0.12) & \textbf{4.93 (0.20)} \\
            \midrule
            \multirow{4}{*}{Graphon 5}
            & 100  & 7.30 (0.01) & 5.36 (0.23) & 9.59 (0.52) & 4.40 (0.15) & \textbf{2.23 (0.12)} \\
            & 200  & 6.31 (0.01) & 4.32 (0.24) & 8.09 (0.38) & 3.81 (0.13) & \textbf{1.76 (0.08)} \\
            & 500  & 3.28 (0.00) & 3.46 (0.25) & 6.62 (0.36) & 3.13 (0.09) & \textbf{1.33 (0.06)} \\
            & 1000 & 3.23 (0.00) & 3.26 (0.19) & 5.72 (0.29) & 2.62 (0.08) & \textbf{1.12 (0.05)} \\
            \bottomrule
        \end{tabular}
    }\label{table:RMSE_node}
\end{table}

As shown in Table \ref{table:RMSE_node}, the estimation errors of all methods decrease monotonically as the node size $n$ increases, which aligns with standard large-sample theory. The proposed MICE method consistently attains the lowest RMSE across all five graphon models, indicating its strong ability to capture heterogeneous topological structures. By adaptively aggregating information from local neighborhoods, MICE effectively leverages larger node scales to further reduce estimation errors and refine the probability estimates.

Table~\ref{table:RMSE_layer} highlights the distinct behaviors of the competing methods as the layer size $K$ grows. Because the ICE algorithm is designed for single-layer networks, its error remains unchanged with respect to $K$; the small fluctuations observed are due to variability in the hyperparameters choices $C_{it}$ and $C_{est}$ across replications. MultiNeSS and DC-MASE show negligible improvement as $K$ increases. Although both models incorporate shared and layer-specific structures, they rely heavily on parameters that are estimated separately for each layer, such as latent positions in MultiNeSS or degree-correction and block matrices in DC-MASE, so increasing $K$ does not provide additional information to refine these layer-specific components. In contrast, both MNS and MICE exhibit substantial gains as $K$ increases. In particular, MICE consistently delivers the highest accuracy across all settings, demonstrating the effectiveness of its iterative strategy in borrowing strength across layers.

\begin{table}[htbp]
    \centering
    \caption{RMSE (×100) across graphons and layer sizes for different methods on a fixed 200-node network. Standard errors are reported in parentheses, and boldface denotes the best performance.}
    \scalebox{0.8}{
        \begin{tabular}{ccccccc}
            \toprule
            Graphon & Layer Size($K$) & ICE & DC-MASE & MultiNeSS & MNS & MICE \\
            \midrule
            \multirow{4}{*}{Graphon 1} 
            & 20  & 4.37 (0.03) & 4.33 (0.33) & 3.13 (0.37) & 4.05 (0.30) &\textbf{1.94 (0.13)} \\
            & 50  & 4.41 (0.02) & 4.12 (0.08) & 2.98 (0.26) & 3.05 (0.14) & \textbf{1.49 (0.10)} \\
            & 100 & 4.39 (0.01) & 4.13 (0.07) & 2.91 (0.19) & 2.57 (0.08) &\textbf{1.24 (0.14)} \\
            & 200 & 4.35 (0.01) & 4.12 (0.06) & 2.88 (0.18) & 2.24 (0.06) & \textbf{1.06 (0.16)} \\
            \midrule
            \multirow{4}{*}{Graphon 2} 
            & 20  & 5.43 (0.01) & 5.89 (0.15) & 4.82 (0.39) & 4.11 (0.20) &\textbf{3.35 (0.17)} \\
            & 50  & 5.51 (0.00) & 5.66 (0.11) & 4.62 (0.27) & 3.67 (0.11) & \textbf{2.57 (0.09)} \\
            & 100 & 5.53 (0.00) & 5.58 (0.11) & 4.57 (0.19) & 3.26 (0.08) & \textbf{2.13 (0.06)} \\
            & 200 & 5.61 (0.00) & 5.51 (0.09) & 4.51 (0.14) & 2.97 (0.08) & \textbf{1.82 (0.08)} \\
            \midrule
            \multirow{4}{*}{Graphon 3}
            & 20  & 5.30 (0.01) & 4.37 (0.10) & 5.22 (0.29) & 3.35 (0.10) & \textbf{2.65 (0.15)} \\
            & 50  & 5.77 (0.01) & 4.31 (0.07) & 4.47 (0.27) & 2.90 (0.05) & \textbf{2.05 (0.10)} \\
            & 100 & 5.71 (0.01) & 4.23 (0.07) & 4.16 (0.21) & 2.60 (0.05) & \textbf{1.72 (0.09)} \\
            & 200 & 5.27 (0.02) & 4.18 (0.26) & 3.99 (0.17) & 2.36 (0.04) &\textbf{1.49 (0.09)} \\
            \midrule
            \multirow{4}{*}{Graphon 4}
            & 20  & 9.61 (0.03) & 8.61 (0.26) & 9.19 (0.28) & 8.43 (0.24) & \textbf{7.32 (0.38)} \\
            & 50  & 9.54 (0.01) & 8.39 (0.18) & 9.13 (0.24) & 7.88 (0.18) & \textbf{6.57 (0.28)} \\
            & 100 & 9.47 (0.01) & 8.37 (0.17) & 9.13 (0.22) & 7.51 (0.15) & \textbf{5.79 (0.29)} \\
            & 200 & 10.10 (0.00) & 8.35 (0.15) & 9.11 (0.19) & 7.14 (0.15) & \textbf{5.29 (0.26)} \\
            \midrule
            \multirow{4}{*}{Graphon 5}
            & 20  & 5.51 (0.02) & 4.36 (0.47) & 7.94 (0.69) & 4.77 (0.32) & \textbf{2.73 (0.20)} \\
            & 50  & 5.43 (0.01) & 4.33 (0.31) & 8.05 (0.56) & 4.21 (0.17) & \textbf{2.07 (0.12)} \\
            & 100 & 6.31 (0.01) & 4.32 (0.24) & 8.09 (0.38) & 3.81 (0.13) &\textbf{1.76 (0.08)} \\
            & 200 & 6.35 (0.01) & 4.24 (0.18) & 8.16 (0.34) & 3.45 (0.09) & \textbf{1.50 (0.06)} \\
            \bottomrule
        \end{tabular}
    }\label{table:RMSE_layer}
\end{table}

\subsection{Parameter Robustness}
\label{sec_simu_para}

\begin{figure}[htbp] 
    \centering
    \includegraphics[width=\textwidth]{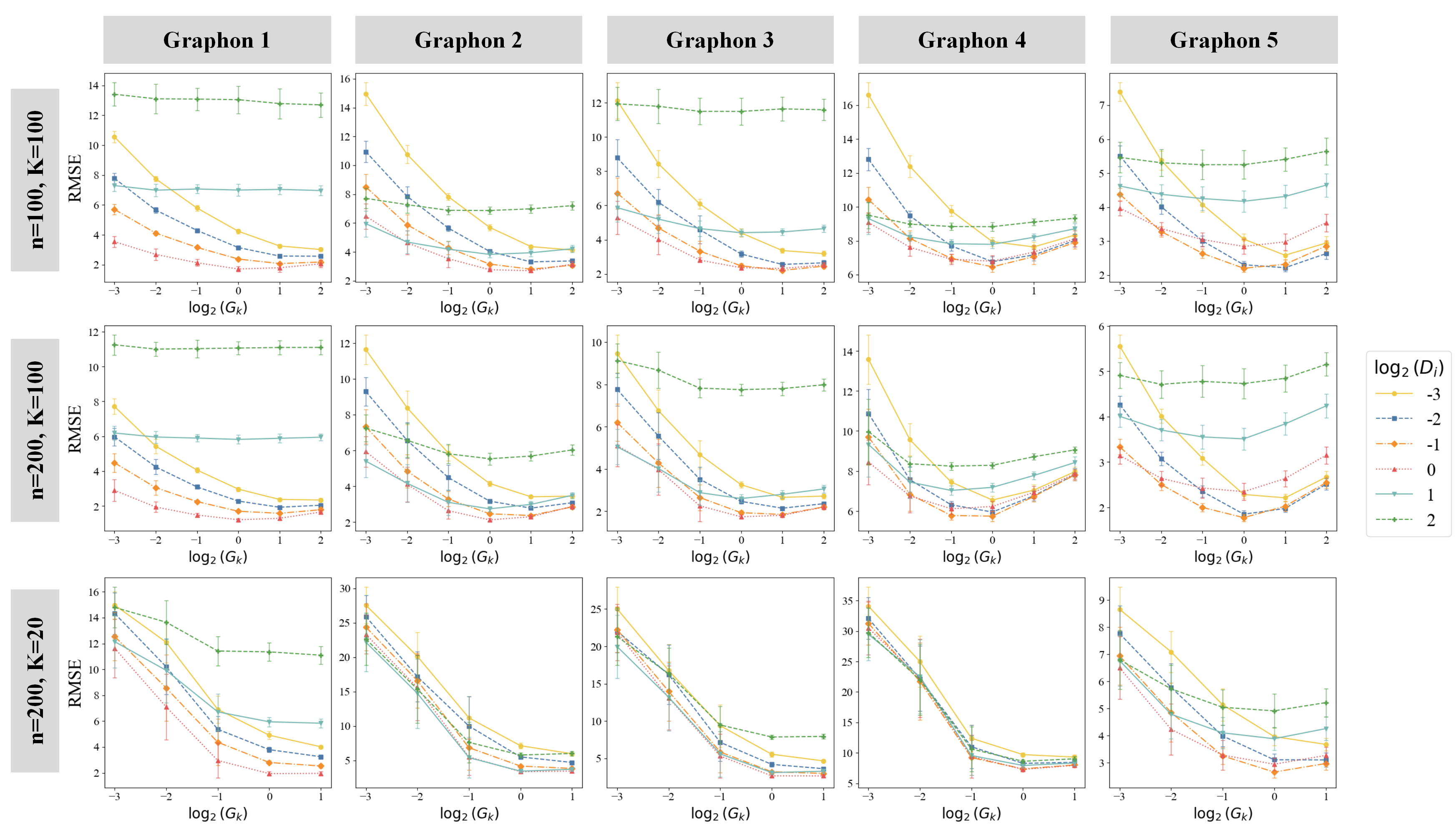} 
    \caption{Average RMSE with standard deviation error bars under varying choices of $D_i$ and $G_k$. (For the setting $(n=200, K=20)$, the case $\log_2 G_k=2$ is omitted because the resulting neighborhood size exceeds the layer size.)} 
    \label{fig:validation} 
\end{figure}

We further examine the sensitivity of the MICE estimator to the neighborhood size scaling constants $D_i$ and $G_k$, as defined in Section \ref{subsec:MICE_alg}. To evaluate robustness with respect to these tuning parameters, we conduct simulation experiments under three representative $(n, K)$ settings: $(100, 100)$, $(200, 100)$, and $(200, 20)$. For each configuration, we vary $D_i$ and $G_k$ over a geometric grid with $\log_2 D_i, \log_2 G_k \in \{-3, -2, -1, 0, 1, 2\}$, corresponding to the explicit values $\{0.125, 0.25, 0.5, 1, 2, 4\}$. For each choice, we compute the RMSE, averaged over 100 independent replications, across all five graphon models. The resulting performance patterns are summarized in Figure~\ref{fig:validation}.

As shown in Figure~\ref{fig:validation}, the results exhibit moderate variation in the optimal choice of tuning parameters. For Graphons 1--3, which feature relatively smooth or monotonic structures, the best performance is obtained using larger neighborhoods ($D_i = G_k = 1$), as this reduces variance without introducing substantial bias. In contrast, for the more complex Graphons 4 and 5, a restricted node-wise neighborhood ($D_i = 0.5$) achieves the lowest RMSE by excluding structurally dissimilar nodes. Because the performance loss from using this restricted node neighborhood in the smoother graphons is minimal, while it consistently improves accuracy for the more irregular ones, we adopt $D_i =0.5$ and $G_k = 1$ as the default configuration in subsequent analyses. Compared with the conventional single-layer bandwidth $D_i = 1$ \citep{zhang2017estimating}, this choice yields a more conservative neighborhood construction that enhances robustness to heterogeneity and irregular structures in multi-layer networks, while maintaining computational efficiency.

\section{Real Data Analysis}
\label{sec:real}

\subsection{Brain Network Data in ADHD Patients}

In this section, we demonstrate the practical utility of the proposed MICE method by analyzing brain network data from patients with Attention-Deficit/Hyperactivity Disorder (ADHD). ADHD is one of the most prevalent childhood-onset neurodevelopmental disorders, and characterizing individual-level heterogeneity in functional brain connectivity is crucial for accurate diagnosis and treatment. We use data from the ADHD-200 Global Competition, which provides demographic information and functional MRI (fMRI) scans for both ADHD patients and typically developing controls (TDC). The data were collected at eight participating institutions and are publicly available at \url{https://www.nitrc.org/frs/?group_id=383}.

To avoid potential site bias, our analysis focuses exclusively on the fMRI data collected at Peking University. This site provides $K=24$ diagnosed ADHD patients for constructing the multi-layer network. For each subject, fMRI time series were recorded from $n=116$ Regions of Interest (ROIs) across $T=172$ time points. The network construction proceeds as follows. First, for each subject $k$ ($k=1, \dots, K$), we compute a sparse correlation matrix between the time series of the $n$ ROIs. Second, to remove weak or spurious correlations, we apply a standard hard-thresholding procedure: an edge is set to 1 whenever the absolute pairwise correlation exceeds 0.7. The resulting adjacency tensor $\mathcal{A}$ contains entries $\mathcal{A}_{ijk} \in \{0,1\}$, indicating the presence or absence of a functional connection between ROI $i$ and ROI $j$ for subject $k$. This process yields a multi-layer network with $K=24$ layers, each of dimension $n \times n$. The adjacency tensor $\cA$ is then used as the input for estimating the underlying connection probability tensor.

Given that the true connection probability tensor $\mathcal{P}$ is unobserved in this dataset, directly evaluating estimation accuracy is not feasible. Following the strategy of \citet{zhang2017estimating}, we assess empirical performance through a link prediction task, which provides a practical proxy for evaluating the estimated probability tensor. Specifically, we generate a partially observed adjacency tensor $\cA_{obs} = \cM \circ \cA$, where $\circ$ denotes the Hadamard product. The mask tensor $\cM$ has independent entries $\cM_{ijk}\sim\text{Bernoulli}(1-\rho)$, so that each entry is masked (set to missing) with probability $\rho$. Based on the estimated probability tensor $\widehat{\cP}$ and a classification threshold $\tau \in (0, 1)$, we then define the True Positive Rate (TPR) and False Positive Rate (FPR) as follows:
\begin{equation}
    \begin{aligned}
        \mathrm{TPR}(\tau) &= 
        \sum_{i,j,k} \mathbb{I}\!\left( \widehat{\cP}_{ijk} > \tau,\, \cA_{ijk} = 1,\, \cM_{ijk} = 0 \right)\Big/
        \sum_{i,j,k} \mathbb{I}\!\left( \cA_{ijk} = 1,\, \cM_{ijk} = 0 \right),\\[6pt]
        \mathrm{FPR}(\tau) &= 
        \sum_{i,j,k} \mathbb{I}\!\left( \widehat{\cP}_{ijk} > \tau,\, \cA_{ijk} = 0,\, \cM_{ijk} = 0 \right)\Big/
        \sum_{i,j,k} \mathbb{I}\!\left( \cA_{ijk} = 0,\, \cM_{ijk} = 0 \right) 
        .
    \end{aligned}
    \label{eq:ROC}
\end{equation}

\begin{figure}[!h]
    \centering
    
    \begin{subfigure}{0.48\textwidth}
        \centering
        \includegraphics[width=\linewidth]{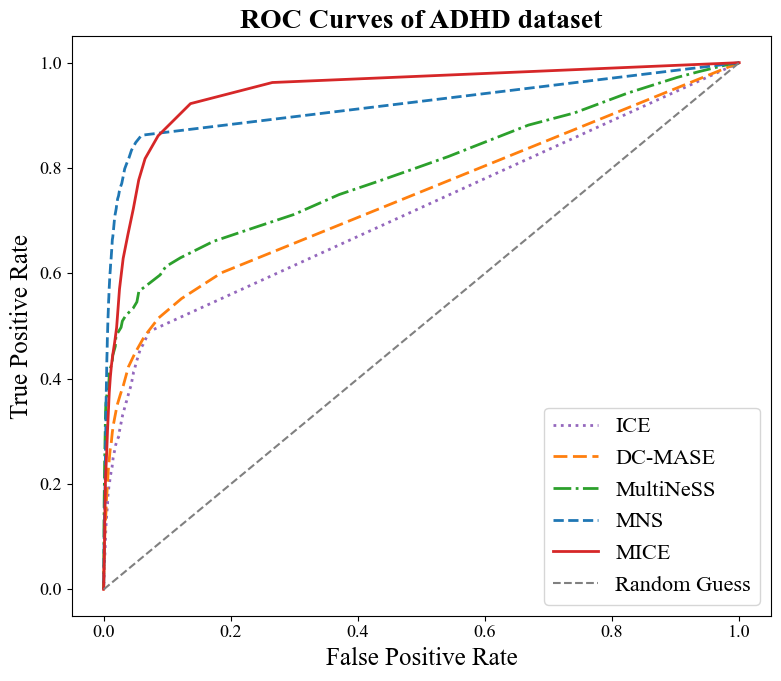}
        \caption{ROC curves for link prediction on the ADHD brain network under a missing rate of $\rho = 0.1$.}
        \label{fig:ROC_ADHD}
    \end{subfigure}
    \hfill
    \begin{subfigure}{0.48\textwidth}
        \centering
        \includegraphics[width=\linewidth]{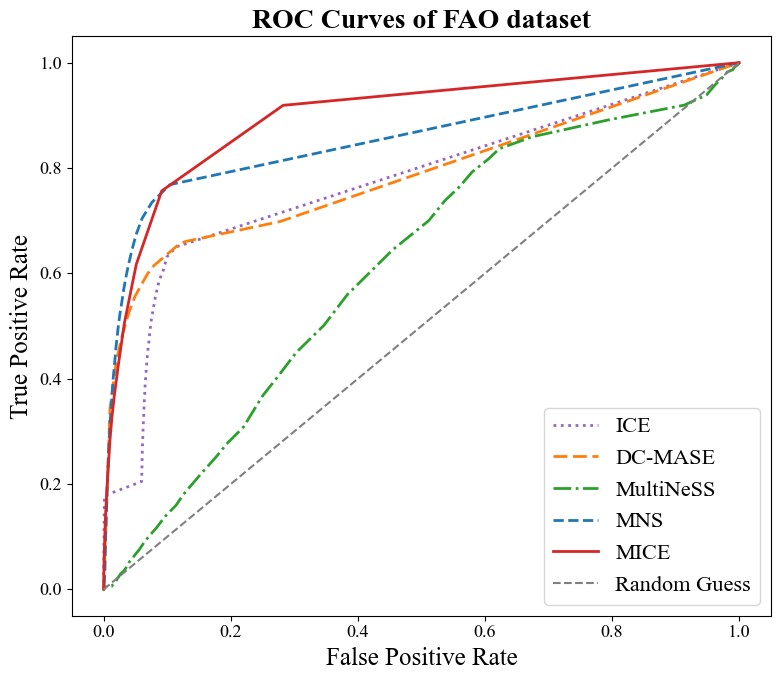}
        \caption{ROC curves for link prediction on the 2023 worldwide food and agricultural trade network under a missing rate of $\rho = 0.2$.}
        \label{fig:ROC_FAO}
    \end{subfigure}
    
    \caption{Receiver operating characteristic curves on two real-world datasets.}
    \label{fig:ab}
\end{figure}

Based on these definitions, we plot the receiver operating characteristic curves for the five comparative methods in Figure \ref{fig:ROC_ADHD} and report their corresponding AUC values in the first row of Table \ref{table:AUC}. Overall, methods designed for multi-layer networks consistently outperform the single-layer benchmark (ICE). Among the multi-layer approaches, the graphon-based estimators MNS and MICE achieve substantially higher predictive accuracy. In particular, the proposed MICE method achieves the highest AUC value of 0.9430, underscoring its robustness and practical utility for analyzing complex real-world datasets.

\subsection{Worldwide Food and Agricultural Trade Network Data}

The worldwide food and agricultural trade network dataset is collected annually by the Food and Agriculture Organization (FAO) of the United Nations, covering the period from 1986 to 2023. The data are publicly available at \url{https://www.fao.org/faostat/en/#data/TM}. In this study, we focus on the most recent trade data from 2023 and extract the reported import quantities for our analysis.

To identify the core nodes of the network, we compute each country's frequency of appearance as both a reporting and a partner nation. Applying a frequency threshold of 1,000 yields $n=137$ key trading countries. This set includes major global economies such as China, Russia, Canada, the United States, Japan, Australia, India, Brazil, and member states of the European Union. To construct a binary multi-layer network, we randomly select $K=100$ distinct item categories. The adjacency tensor is defined using trade volumes: a strong connection between country $i$ and country $j$ in item category $k$ is recorded (that is, $\mathcal{A}_{ijk} = 1$) if the traded quantity exceeds 1,000 units; otherwise, $\mathcal{A}_{ijk} = 0$.

We adopt a validation strategy similar to that used in the ADHD brain network analysis to evaluate the performance of the MICE method. In this experiment, we set the missing rate of the mask tensor to $\rho = 0.2$. Figure \ref{fig:ROC_FAO} displays the ROC curves for each method, and the corresponding AUC values are reported in the second row of Table \ref{table:AUC}. Consistent with our earlier findings, MICE attains the highest AUC value of 0.8976, further demonstrating the robustness and practical effectiveness of our approach in real-world applications.

\begin{table}[htbp]
    \centering
    \caption{Performance on real-world datasets: AUC values for link prediction on ADHD and FAO (first two rows), and prediction precision for forecasting the 2023 FAO network based on 2022 data (last row).}
    \begin{tabular}{lccccc}
        \toprule
        Dataset & ICE & DC-MASE & MultiNeSS & MNS & MICE \\
        \midrule
        ADHD & 0.7159 & 0.7422 & 0.7929 & 0.9183 & \textbf{0.9430} \\
        FAO  & 0.7720 & 0.7842 & 0.6123 & 0.8547 & \textbf{0.8976} \\
        FAO-prediction & 20.13\% &15.09\% & 19.50\% & 4.30\% & \textbf{23.16\%}\\
        \bottomrule
    \end{tabular}
    \label{table:AUC}
\end{table}

To further assess the predictive capability of our model, we conduct a temporal link prediction experiment that uses the 2022 trade network to forecast the 2023 network structure. Given the relative stability of annual trade volumes, we adjust the edge-definition criterion: a connection is considered present if the trade volume exceeds 100 units. We randomly sample $K=200$ network layers and apply our method to the 2022 data. The validation task focuses on identifying emerging links. A prediction is counted as successful if the model assigns a high probability to node pairs that are unconnected in the 2022 network (trade volume $\le$ 100) but form a connection in the corresponding 2023 network (trade volume $>$ 100).

We apply all five methods to the 2022 data to generate the probability tensor $\widehat{\cP}^{(2022)}$. A potential link is predicted whenever the estimated probability exceeds 0.5. To evaluate performance specifically in identifying emerging connections, we restrict attention to node-item triples $(i, j, k)$ for which no link existed in 2022 ($\cA_{ijk}^{(2022)}=0$). A prediction is counted as correct if the model successfully forecasts the presence of a link in the 2023 network ($\cA_{ijk}^{(2023)}=1$). The prediction precision is computed as:
$$
\text{Precision}
=
\frac{
\sum_{i,j,k}
\mathbb{I}\left(
\widehat{\mathcal{P}}_{ijk}^{(2022)} > 0.5,\,
\mathcal{A}_{ijk}^{(2023)} = 1,\,
\mathcal{A}_{ijk}^{(2022)} = 0
\right)
}{
\sum_{i,j,k}
\mathbb{I}\left(
\widehat{\mathcal{P}}_{ijk}^{(2022)} > 0.5,\,
\mathcal{A}_{ijk}^{(2022)} = 0
\right)
}.
$$
The prediction results for all five models are presented in the third row of Table \ref{table:AUC}.

Among all five methods, MICE achieves the highest precision. However, the overall precision levels across methods remain limited. This suggests that relying solely on structural information from past trade networks, without incorporating additional covariates, may be insufficient for forecasting complex future trade patterns.

\begin{figure}[!h]
    \centering
    \begin{subfigure}{0.48\textwidth}
        \centering
        \includegraphics[width=\linewidth]{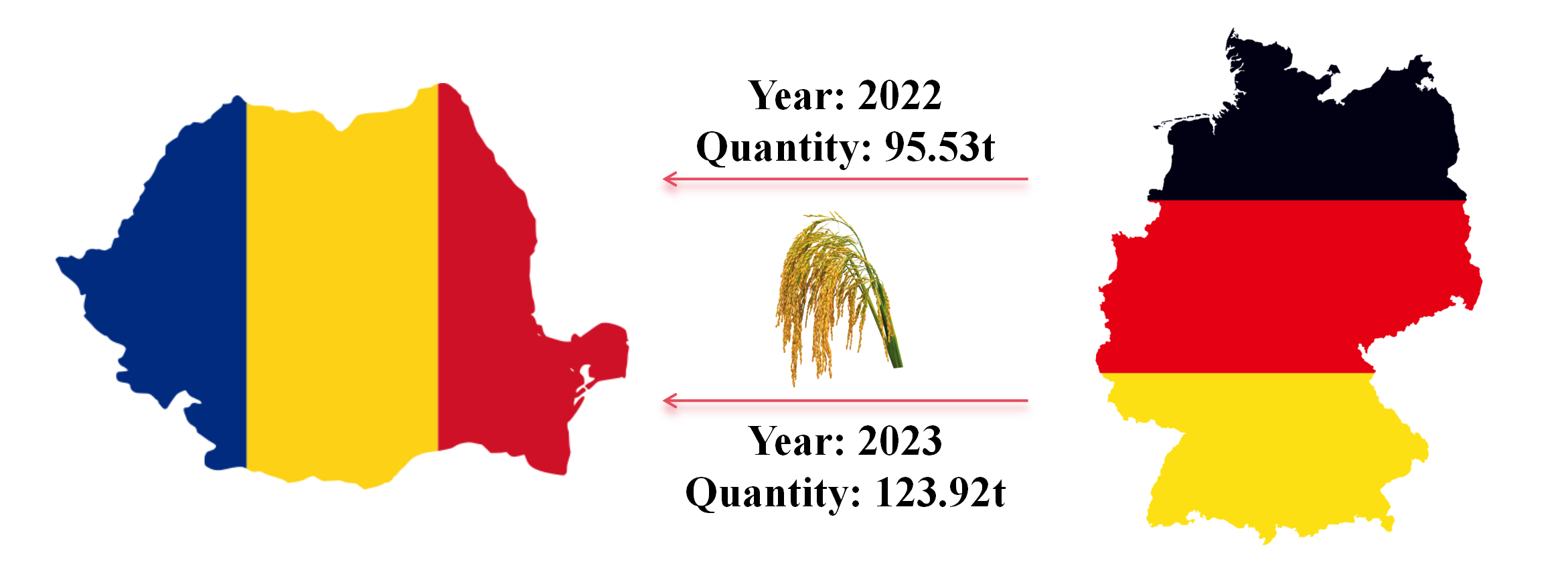}
    \end{subfigure}
    \hfill
    \begin{subfigure}{0.48\textwidth}
        \centering
        \includegraphics[width=\linewidth]{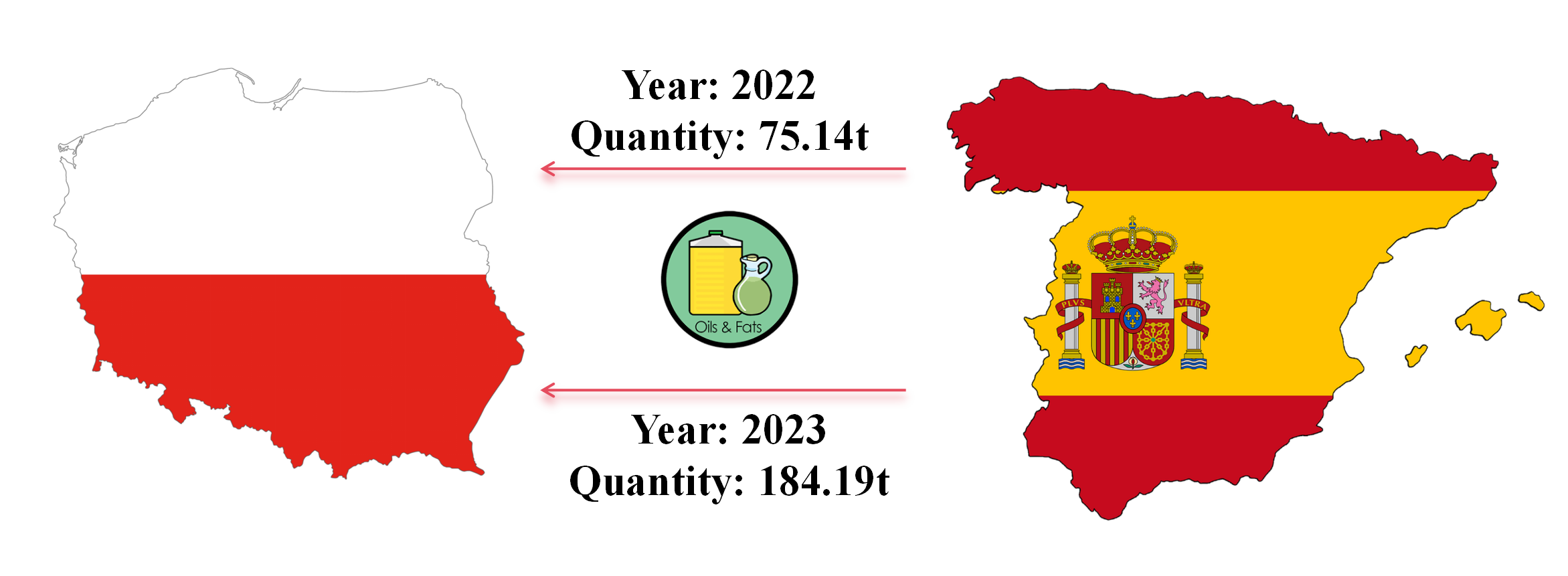}
    \end{subfigure}
    \caption{Two examples of validated link predictions. Emerging connections in 2023 are correctly forecasted from 2022 data with estimated probabilities of 0.712 and 0.626.}
    \label{fig:import_change}
\end{figure}

To illustrate the predictive capability of the MICE method, we highlight two representative instances shown in Figure \ref{fig:import_change}, where the model successfully identifies latent trade potential. In both examples, the method assigns high probabilities to country-item pairs that are not yet connected in 2022 (volume below $100$ tons) but become connected in 2023. The first case involves Item Code 30 (Rice, paddy). In 2022, Romania imported 95.53 tons from Germany, just below the threshold ($\mathcal{A}_{ijk}^{(2022)} = 0$). Nevertheless, MICE assigns a probability of 0.712 to this prospective connection. This prediction is validated in 2023, when the trade volume rises to 123.92 tons and the link becomes active. A similar pattern is observed for Item Code 1274 (Chemically modified fats and oils). Poland imported 75.14 tons from Spain in 2022, which was insufficient to form a link, yet the model predicts a probability of 0.626. By 2023, the volume increases sharply to 184.19 tons, confirming a strong trade relationship. These examples illustrate the method's ability to detect emerging links ahead of time.

\section{Conclusion}
\label{sec:conclusion}

This study introduces an iterative estimation method for connection probabilities in multi-layer networks that progressively updates neighboring sets and probability estimates across both layers and nodes. The proposed method is conceptually intuitive, straightforward to implement, and computationally efficient. Focusing on multi-layer undirected binary networks, we establish the theoretical consistency of the estimator. Extensive numerical experiments show that our approach consistently outperforms existing methods across diverse graphon functions and network scales. Moreover, the method demonstrates strong performance in link prediction tasks on real-world datasets, underscoring its practical usefulness.

Despite these contributions, several directions remain open for future research. Although we establish theoretical consistency, achieving the optimal convergence rate requires a large number of layers, which may be difficult in practice. In addition, the current framework assumes a common node set across all layers; relaxing this assumption to allow for varying or partially overlapping node sets would substantially broaden the method's applicability to heterogeneous network settings. Finally, the current approach employs a uniform quantile threshold to determine neighborhood sizes. Developing data-driven, adaptive mechanisms that assign neighborhood sizes based on local structural density represents a promising avenue for further improving estimation accuracy.

\section*{Supplementary Materials}

Supplementary material available at Statistica Sinica online includes additional simulation results and detailed proofs of main theorems and supporting lemmas.
\par
\section*{Acknowledgements}
This research is supported by NSF China (12401370), the Summit Advancement Disciplines of Zhejiang Province (Zhejiang Gongshang University - Statistics) and Collaborative Innovation Center of Statistical Data Engineering Technology \& Application.
\par


\bibhang=1.7pc
\bibsep=2pt
\fontsize{9}{14pt plus.8pt minus .6pt}\selectfont
\renewcommand\bibname{\large \bf References}
\expandafter\ifx\csname
natexlab\endcsname\relax\def\natexlab#1{#1}\fi
\expandafter\ifx\csname url\endcsname\relax
\def\url#1{\texttt{#1}}\fi
\expandafter\ifx\csname urlprefix\endcsname\relax\def\urlprefix{URL}\fi

{
    \bibliographystyle{chicago}
    \bibliography{references}

@article{chandna2020nonparametric,
  title={Nonparametric regression for multiple heterogeneous networks},
  author={Chandna, Swati and Maugis, Pierre-Andre},
  journal={arXiv preprint arXiv:2001.04938},
  year={2020}
}

@inproceedings{ICE,
 author = {Qin, Yichen and Yu, Linhan and Li, Yang},
 booktitle = {Advances in Neural Information Processing Systems},
 pages = {1155--1166},
 title = {Iterative Connecting Probability Estimation for Networks},
 volume = {34},
 year = {2021}
}

@article{zhang2017estimating,
  title={Estimating network edge probabilities by neighbourhood smoothing},
  author={Zhang, Yuan and Levina, Elizaveta and Zhu, Ji},
  journal={Biometrika},
  volume={104},
  number={4},
  pages={771--783},
  year={2017},
  publisher={Oxford University Press}
}

@article{macdonald2022latent,
  title={Latent space models for multiplex networks with shared structure},
  author={MacDonald, Peter W and Levina, Elizaveta and Zhu, Ji},
  journal={Biometrika},
  volume={109},
  number={3},
  pages={683--706},
  year={2022},
  publisher={Oxford University Press}
}

@article{agterberg2025joint,
  title={Joint spectral clustering in multilayer degree-corrected stochastic blockmodels},
  author={Agterberg, Joshua and Lubberts, Zachary and Arroyo, Jes{\'u}s},
  journal={Journal of the American Statistical Association},
  volume = {120},
  number = {551},
  pages = {1607--1620},
  year = {2025},
  publisher={Taylor \& Francis}
}

@article{kivela2014multilayer,
  title={Multilayer networks},
  author={Kivel{\"a}, Mikko and Arenas, Alex and Barthelemy, Marc and Gleeson, James P and Moreno, Yamir and Porter, Mason A},
  journal={Journal of complex networks},
  volume={2},
  number={3},
  pages={203--271},
  year={2014},
  publisher={Oxford University Press}
}

@ARTICLE{Pham2022,
  author={Pham, Phu and Nguyen, Loan T. T. and Nguyen, Ngoc Thanh and Pedrycz, Witold and Yun, Unil and Vo, Bay},
  journal={IEEE Transactions on Systems, Man, and Cybernetics: Systems}, 
  title={ComGCN: Community-Driven Graph Convolutional Network for Link Prediction in Dynamic Networks}, 
  year={2022},
  volume={52},
  number={9},
  pages={5481-5493},
  doi={10.1109/TSMC.2021.3130149}}

@article{Xu03072023,
author = {Shirong Xu and Yaoming Zhen and Junhui Wang},
title = {Covariate-Assisted Community Detection in Multi-Layer Networks},
journal = {Journal of Business \& Economic Statistics},
volume = {41},
number = {3},
pages = {915--926},
year = {2023},
publisher = {ASA Website},
doi = {10.1080/07350015.2022.2085726},
URL = {    
        https://doi.org/10.1080/07350015.2022.2085726
},
eprint = {    
        https://doi.org/10.1080/07350015.2022.2085726
}
}

@article{wolfe2013graphon,
  title={Nonparametric graphon estimation},
  author={Wolfe, Patrick J and Olhede, Sofia C},
  journal={arXiv preprint arXiv:1309.5936},
  year={2013}
}

@article{Gao2015graphon,
author = {Chao Gao and Yu Lu and Harrison H. Zhou},
title = {{Rate-optimal graphon estimation}},
volume = {43},
journal = {The Annals of Statistics},
number = {6},
publisher = {Institute of Mathematical Statistics},
pages = {2624 -- 2652},
year = {2015},
doi = {10.1214/15-AOS1354},
URL = {https://doi.org/10.1214/15-AOS1354}
}

@article{Choi2017graphon,
author = {David Choi},
title = {{Co-clustering of nonsmooth graphons}},
volume = {45},
journal = {The Annals of Statistics},
number = {4},
publisher = {Institute of Mathematical Statistics},
pages = {1488 -- 1515},
year = {2017},
doi = {10.1214/16-AOS1497},
URL = {https://doi.org/10.1214/16-AOS1497}
}

@article{Chandna2021graphon,
    author = {Chandna, S and Olhede, S C and Wolfe, P J},
    title = {Local linear graphon estimation using covariates},
    journal = {Biometrika},
    volume = {109},
    number = {3},
    pages = {721-734},
    year = {2022},
}

@article{dynamicPensky,
author = {Marianna Pensky},
title = {{Dynamic network models and graphon estimation}},
volume = {47},
journal = {The Annals of Statistics},
number = {4},
publisher = {Institute of Mathematical Statistics},
pages = {2378 -- 2403},
year = {2019},
doi = {10.1214/18-AOS1751},
URL = {https://doi.org/10.1214/18-AOS1751}
}

@article{wang2021optimal,
  title={Optimal change point detection and localization in sparse dynamic networks},
  author={Wang, Daren and Yu, Yi and Rinaldo, Alessandro},
  journal={The Annals of Statistics},
  volume={49},
  number={1},
  pages={203--232},
  year={2021},
  publisher={JSTOR}
}

@article{MSBM_Paul,
author = {Subhadeep Paul and Yuguo Chen},
title = {{Consistent community detection in multi-relational data through restricted multi-layer stochastic blockmodel}},
volume = {10},
journal = {Electronic Journal of Statistics},
number = {2},
publisher = {Institute of Mathematical Statistics and Bernoulli Society},
pages = {3807 -- 3870},
year = {2016},
doi = {10.1214/16-EJS1211},
URL = {https://doi.org/10.1214/16-EJS1211}
}

@article{aldous1981representations,
  title={Representations for partially exchangeable arrays of random variables},
  author={Aldous, David J},
  journal={Journal of Multivariate Analysis},
  volume={11},
  number={4},
  pages={581--598},
  year={1981},
  publisher={Elsevier}
}

@article{hoover1979relations,
  title={Relations on Probability Spaces and Arrays of Random Variables},
  author={Hoover, Douglas N},
  journal={Preprint, Institute for Advanced Study},
  year={1979}
}

@article{rahiminejad2019topological,
  title={Topological and functional comparison of community detection algorithms in biological networks},
  author={Rahiminejad, Sara and Maurya, Mano R and Subramaniam, Shankar},
  journal={BMC bioinformatics},
  volume={20},
  number={1},
  pages={212},
  year={2019},
  publisher={Springer}
}

@article{calderer2021community,
  title={Community detection in large-scale bipartite biological networks},
  author={Calderer, Gen{\'\i}s and Kuijjer, Marieke L},
  journal={Frontiers in Genetics},
  volume={12},
  pages={649440},
  year={2021},
  publisher={Frontiers Media SA}
}

@article{eagle2009inferring,
  title={Inferring friendship network structure by using mobile phone data},
  author={Eagle, Nathan and Pentland, Alex and Lazer, David},
  journal={Proceedings of the national academy of sciences},
  volume={106},
  number={36},
  pages={15274--15278},
  year={2009},
  publisher={National Academy of Sciences}
}

@inproceedings{leskovec2010empirical,
  title={Empirical comparison of algorithms for network community detection},
  author={Leskovec, Jure and Lang, Kevin J and Mahoney, Michael},
  booktitle={Proceedings of the 19th international conference on World wide web},
  pages={631--640},
  year={2010}
}

@article{boccaletti2006complex,
  title={Complex networks: Structure and dynamics},
  author={Boccaletti, Stefano and Latora, Vito and Moreno, Yamir and Chavez, Martin and Hwang, D-U},
  journal={Physics reports},
  volume={424},
  number={4-5},
  pages={175--308},
  year={2006},
  publisher={Elsevier}
}

@article{goldenberg2010survey,
  title={A survey of statistical network models},
  author={Goldenberg, Anna and Zheng, Alice X and Fienberg, Stephen E and Airoldi, Edoardo M},
  journal={Foundations and Trends{\textregistered} in Machine Learning},
  volume={2},
  number={2},
  pages={129--233},
  year={2010},
  publisher={Emerald Publishing Limited}
}

@article{he2026jointestimationedgeprobabilities,
      title={Joint Estimation of Edge Probabilities for Multi-layer Networks via Neighborhood Smoothing}, 
      author={Yong He and Zizhou Huang and Bingyi Jing and Diqing Li},
      journal={arXiv preprint arXiv:2601.20219},
      year={2026},
}

@InProceedings{pmlr-v32-chan14,
  title = 	 {A Consistent Histogram Estimator for Exchangeable Graph Models},
  author = 	 {Chan, Stanley and Airoldi, Edoardo},
  booktitle = 	 {Proceedings of the 31st International Conference on Machine Learning},
  pages = 	 {208--216},
  year = 	 {2014},
  volume = 	 {32},
  number =       {1},
  series = 	 {Proceedings of Machine Learning Research},
}

@article{Bickel2009ANV,
  title={A nonparametric view of network models and Newman–Girvan and other modularities},
  author={Peter J. Bickel and Aiyou Chen},
  journal={Proceedings of the National Academy of Sciences},
  year={2009},
  volume={106},
  pages={21068 - 21073}
}

@article{Jing2021community,
author = {Jing, Bingyi and Li, Ting and Lyu, Zhongyuan and Xia, Dong},
title = {Community detection on mixture multilayer networks via regularized tensor decomposition},
volume = {49},
journal = {The Annals of Statistics},
number = {6},
publisher = {Institute of Mathematical Statistics},
pages = {3181 -- 3205},
year = {2021}
}

@article{zhao2019change,
      title={Change-point detection in dynamic networks via graphon estimation}, 
      author={Zifeng Zhao and Li Chen and Lizhen Lin},
      journal={arXiv preprint arXiv:1908.01823},
      year={2019},
}
}

\vskip .65cm
\noindent
\textbf{Dingzi Guo}, Institute for Financial Studies, Shandong University, Jinan, China
\vskip 2pt
\noindent
E-mail: guodingzi@mail.sdu.edu.cn
\vskip 2pt

\noindent
\textbf{Corresponding author: Diqing Li}, School of Statistics and Data Science, Zhejiang Gongshang University, Hangzhou, China
\vskip 2pt
\noindent
E-mail: dqli@mail.zjgsu.edu.cn

\noindent
\textbf{Jingyi Wang}, Institute for Financial Studies, Shandong University, Jinan, China
\vskip 2pt
\noindent
E-mail: 202432059@mail.sdu.edu.cn

\noindent
\textbf{Wen-Xin Zhou},  Department of Information \& Decision Sciences, College of Business Administration, University of Illinois Chicago, Chicago, USA
\vskip 2pt
\noindent
E-mail: wenxinz@uic.edu

\end{document}